\documentclass[lettersize,journal]{IEEEtran}
\usepackage{amsmath,amsfonts}
\usepackage{algorithmic}
\usepackage{algorithm}
\usepackage{array}
\usepackage[caption=false,font=normalsize,labelfont=sf,textfont=sf]{subfig}
\usepackage{textcomp}
\usepackage{stfloats}
\usepackage{url}
\usepackage{verbatim}
\usepackage{graphicx}
\usepackage{cite}
\usepackage{amsmath}
\begin{document}

\title{Trainable Proximal Gradient Descent Based Channel Estimation for mmWave Massive MIMO Systems}

\author{Peicong Zheng, Xuantao Lyu, and Yi Gong~\IEEEmembership{Senior Member,~IEEE}
\thanks{P. Zheng and Y. Gong are with the Department of Electrical and	Electronic Engineering, Southern University of Science and Technology,	Shenzhen, China (e-mail: 12131052@mail.sustech.edu.cn; gongy@sustech.edu.cn). P. Zheng is also with Peng Cheng Laboratory, Shenzhen, China. X. Lyu is with  the same institute (e-mail: lvxt@pcl.ac.cn).}
}



\maketitle 

\begin{abstract}
In this letter, we address the problem of millimeter-Wave channel estimation in massive MIMO communication systems.
Leveraging the sparsity of the mmWave channel in the beamspace, we formulate the estimation problem as a sparse signal recovery problem.
To this end, we propose a deep learning based trainable proximal gradient descent network (TPGD-Net).
The TPGD-Net unfolds the iterative proximal gradient descent (PGD) algorithm into a layer-wise network, with the gradient descent step size set as a trainable parameter. 
Additionally, we replace the proximal operator in the PGD algorithm with a neural network that exploits data-driven prior channel information to perform the proximal operation implicitly.
To further enhance the transfer of feature information across layers, we introduce the cross-layer feature attention fusion module into the TPGD-Net.
Our simulation results on the Saleh-Valenzuela channel model and the DeepMIMO dataset demonstrate the superior performance of TPGD-Net compared to state-of-the-art mmWave channel estimators.

\end{abstract}


\section{Introduction}
Millimeter wave (mmWave) massive multiple-input-multiple-output (massive MIMO) has been recognized as a key enabling technology for the 5G and beyond 5G communication systems\cite{hong2021role}.
It has shown its great potential to significantly improve the spectral efficiency and increase the system capacity owning to the abundance of available bandwidth at mmWave frequency and the high beamforming gain attained from large-scale antenna array.
However, the conventional multi-antenna configuration, employing one dedicated radio frequency (RF) chain per antenna element, is impractical for mmWave massive MIMO system from the perspectives of both hardware cost and power consumption. 
Therefore, the use of lens antenna arrays in mmWave massive MIMO systems has attracted considerable interest due to its potential to reduce the cost of the RF chain while maintaining acceptable performance\cite{zeng2016millimeter}.
Specifically, a lens antenna array consists of an electromagnetic (EM) lens and an antenna array with elements positioned in the focal region of the EM lens.
The EM lens focuses incoming mmWave signals from various directions onto the antenna array, concentrating the signal power to a sub-region of the antenna array.
This allows a subset of antenna elements to be activated, reducing both the cost of the RF chain and the complexity of signal processing.

Acquiring accurate channel state information is crucial to realize the potential benefits of mmWave massive MIMO systems with lens antenna arrays.
However, the limited number of RF chains causes a reduction in the dimensionality of the received signal, which in turn, poses a significant challenge to channel estimation.
Due to the limited scattering characteristics of mmWave propagation and the power focusing capability of the lens antenna array, the spatial domain channel can be transformed into its sparse beamspace representation.
As a result, the estimation problem can be further formulated as a sparse signal recovery problem and solved by employing compressed sensing (CS) techniques such as orthogonal matching pursuit (OMP)\cite{alkhateeb2014channel} and approximate message passing (AMP)\cite{donoho2009message}.
However, the above CS-based sparse channel estimation method only captures the sparsity of the channel, resulting in degraded estimation performance\cite{yang2018beamspace}.

Recently, there has been a growing interest in combining model-based compressed sensing with deep learning to improve beamspace channel estimation.
In \cite{he2018deep}, He \textit{et al.} unfold the AMP algorithm into a learnable network, where the shrinkage function is replaced by a denoising convolutional neural network. 
In \cite{wei2019amp}, Wei \textit{et al.} use a deep residual learning network to refine the coarse estimate obtained from the unfolded AMP network.
To further exploit the prior information of beamspace channel, in \cite{wei2020deep}, Wei \textit{et al.} propose a prior-aided Gaussian mixture learned approximate message passing (GM-LAMP) by incorporating the unfolded AMP network with a shrinkage function derived from the Gaussian mixture model of beamspace channel.
While deep unfolding-based networks have shown promising simulation performance with low computational cost, they rely on prior information that is not always consistent with realistic channel models, and in some cases, these networks may fail to converge\cite{rangan2019convergence,gao2022data}.
In addition, since iterative model-based methods output estimated channel at each iteration, corresponding deep unfolding networks must take channel as input and output of each layer, resulting in feature-to-channel information loss at each layer and hindering the transfer of feature information across layers.
Overall, the performance of these deep unfolding-based networks is limited by the representation ability of prior information and the bottleneck of feature information transmission inherent in the network structure.
To address the aforementioned issues, we propose a trainable proximal gradient descent network (TPGD-Net) for mmWave channel estimation.
The TPGD-Net is designed to unfold the iterative proximal gradient descent (PGD) algorithm into a layer-wise network, with the step size in the gradient descent module of PGD set as a trainable parameter to accelerate the convergence.
The proximal mapping module of PGD is replaced by a tailored neural network to extract data-driven prior information of mmWave channels and perform proximal operation implicitly.  
We introduce the cross-layer feature attention fusion (CLFAF) module into TPGD-Net to address the intrinsic information loss issue.
We compare TPGD-Net with both CS-based methods and existing deep unfolding-based channel estimation methods. 
Simulation results show that the proposed TPGD-Net outperforms existing algorithms under the Saleh-Valenzuela channel model and the DeepMIMO dataset \cite{Alkhateeb2019}.
Furthermore, the TPGD-Net provides comparable or superior performance compared to GM-LAMP while reducing the required number of RF chains by half.
\section{SYSTEM MODEL}
In this section, we first introduce the mmWave massive MIMO system with lens antenna array and formulate the channel estimation as a signal recovery problem.
\begin{figure}[!t]
	\centering
	\includegraphics[scale=0.40]{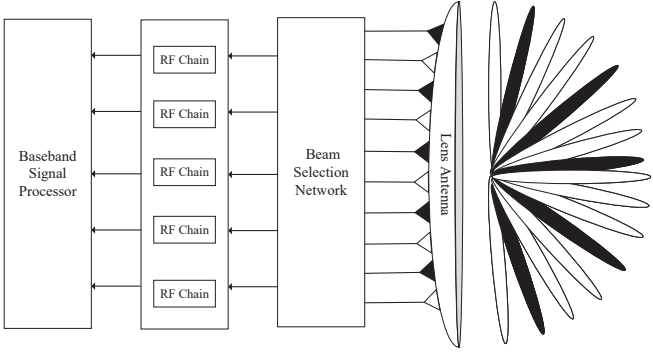}
	\caption{MmWave massive MIMO system with a lens antenna array.}
	\label{fig_1}
\end{figure}
Fig. 1 depicts a mmWave massive MIMO system in which a base station (BS) utilizes a lens antenna array with $N$ antennas and $N_{RF}$ RF chains to serve $K$ single-antenna users.
The mmWave channels between BS and users are characterized by the Saleh-Valenzuela(SV) channel model\cite{9685707}.
Specifically, the uplink channel $\hat{\mathbf{h}}_{k} \in \mathbb{C}^{N\times 1}$ of the $k$th user is given by:
\begin{equation}
	\hat{\mathbf{h}}_{k}=\sqrt{\frac{N}{L_k}} \sum_{l=1}^{L_k} \beta_{k}^{l} \mathbf{a}\left(\theta_{k}^{l}\right),
	\label{eq1}
\end{equation}
where ${L_k}$ denotes the number of resolvable paths, $\beta_{k}^{l}$ is the gain of the $l$th path, and $\theta_{k}^{l}$ denotes the spatial direction of the $l$th path. The spatial direction $\theta_{k}^{l}$  is defined as ${d} \sin \phi_{k}^{l} /{\lambda}$, where $\lambda$ is the wavelength of the carrier, $\phi_{k}^{l}$ is the physical direction, and $d$ denotes the antenna spacing satisfying $d=\lambda / 2$.
For the uniform linear array, the steering vector is expressed as:
\begin{equation}
	\mathbf{a}(\theta)=\frac{1}{\sqrt{N}}\left[1,e^{-j 2 \pi \theta}, \ldots, e^{-j 2 \pi\left(N-1\right) \theta}\right]^{T}.
\end{equation}

Essentially, the lens antenna array is modeled as a spatial discrete Fourier transform matrix $\mathbf{U} \in \mathbb{C}^{N \times N}$, which converts the spatial channel into a beamspace channel. 
Therefore, the beamspace representation of $\mathbf{h}_k$ can be expressed as ${\mathbf{h}}_k=\mathbf{U} \hat{\mathbf{h}}_{k},$
where $\mathbf{U}$ is a set of orthogonal array steering vectors and is defined as $\mathbf{U}=\left[\mathbf{a}\left(\bar{\theta}_1\right), \mathbf{a}\left(\bar{\theta}_2\right), \cdots,
\mathbf{a}\left(\bar{\theta}_N\right)\right]$.
Here, $\bar{\theta}_n=\left(n-\left({N+1}\right)/{2}\right)/N$ represents the spatial directions defined by the lens antenna.

In this letter, we focus on the pilot-assisted uplink channel estimation for mmWave massive MIMO systems.
We assume that appropriate orthogonal pilot assignment has been performed before the uplink channel estimation phase.
Due to the orthogonality of the pilots, the channel estimation for each user can be performed independently.
During the estimation phase, the $k$th user transmits a pilot sequence, $\mathbf{s}_k=\left[s_{k,1},s_{k,2},\cdots,s_{k,M}\right]$.
The received signal at the BS at time instant $m$ is denoted by ${{\mathbf{y}}}_{k,m}\in \mathbb{C}^{N_{RF}\times 1}$ and expressed as:
\begin{equation}
	{{\mathbf{y}}}_{k,m}=\mathbf{W}_{m} {\mathbf{h}}_k s_{k,m}+\mathbf{W}_{m} {\mathbf{n}}_{k,m},
\end{equation}
where $\mathbf{W}_{m}\in \mathbb{C}^{N_{RF}\times N}$ is the beam selection matrix and $\mathbf{n}_{k,m} \sim \mathcal{C N}\left(0, \sigma_n^2 \mathbf{I}_{\mathbf{N}}\right)$ is the Gaussian noise vector with the noise power $\sigma_n^2$. 
To facilitate the explanation, the value of $s_{k,m}$ is set to $1$ as the pilot symbols are known to the BS.
This simplification can be generalized to other pilot design schemes without loss of generality.
After receiving the pilot sequence $\mathbf{s}_k$, the received signal sequence ${\mathbf{y}}_k$ is equivalently rewritten as follows:
\begin{equation}
	{\mathbf{y}}_k=\left[\mathbf{y}_{k, 1}^T,\mathbf{y}_{k, 2}^T \ldots, \mathbf{y}_{k, M}^T\right]^T=\hat{\mathbf{W}} {\mathbf{h}}_k+\hat{\mathbf{W}} \mathbf{n}_k,
\end{equation}
where $\hat{\mathbf{W}}=\left[\mathbf{W}_1^T, \mathbf{W}_2^T, \ldots, \mathbf{W}_M^T\right]^T$ is the stacked beam selection matrix, and $\mathbf{n}_k=\left[\mathbf{n}_{k,1}^T, \mathbf{n}_{k,2}^T,\ldots, \mathbf{n}_{k,M}^T\right]^T.$
The received signal at the BS of all $K$ users is given as:
\begin{equation}
\mathbf{Y}=\mathbf{W H}+\mathbf{W N},
\label{eq7}
\end{equation}
where $\mathbf{Y}=\left[{\mathbf{y}}_1^T,{\mathbf{y}}_2^T, \ldots, {\mathbf{y}}_K^T\right]^T$, $\mathbf{H}=\left[{\mathbf{h}}_1^T, {\mathbf{h}}_2^T, \ldots, {\mathbf{h}}_K^T\right]^T$, and  $\mathbf{N}=\left[\mathbf{n}_1^T,\mathbf{n}_2^T, \ldots,\mathbf{n}_K^T\right]^T$, and the equivalent beam selection matrix $\mathbf{W}$ is defined as the Kronecker product of the identity matrix and the matrix $\hat{\mathbf{W}}$.
Therefore, the problem of estimating the mmWave channels is simplified to estimating the beamspace channel $\mathbf{H}$, given the received signal $\mathbf{Y}$ and the equivalent beam selection matrix $\mathbf{W}$.
\section{Methodology}
In this section, we first provide a brief overview of the proximal gradient descent algorithm for beamspace channel estimation. 
Then, we present our proposed deep learning-based trainable proximal gradient descent network.
Given the sparsity of the beamspace mmWave channel, channel estimation in (\ref{eq7}) can be viewed as a sparse signal recovery problem and solved using the Bayesian Maximum A Posteriori (MAP) framework\cite{9685707}. 
Specifically, the sparse beamspace channel $\mathbf{H}$ can be estimated by solving the following optimization problem:
\begin{equation}
    \underset{\mathbf{H}}{\operatorname{min}}   
\|\mathbf{Y}-\mathbf{W H}\|_2^2 /2+\lambda \psi(\mathbf{H}),
	\label{eq8}
\end{equation}
where $\|\mathbf{Y}-\mathbf{W H}\|_2^2 /2 $ is the data fidelity terms that ensures consistency between the estimated channel and the observed measurements.
The regularization term $\psi(\mathbf{H})$ incorporates channel prior information to regularize the solution of (\ref{eq8}), and the weighted parameter $\lambda \geq 0$ balances the data-fidelity and regularization term.

One of the effective methods for solving (\ref{eq8}) is the Proximal Gradient Descent algorithm \cite{beck2009fast}, which generates the estimated channel by iteratively updating the following steps: 
\begin{align}
	\mathbf{V}_{t}&=\mathbf{H}_{t-1}+\alpha \mathbf{W}^{{H}}\left(\mathbf{Y}-\mathbf{W H}_{t-1}\right),\label{eq9}\\
	\mathbf{H}_{t}&=\underset{\mathbf{H}}{\arg \min } \frac{1}{2}\left\|\mathbf{H}-\mathbf{V}_{t}\right\|_2^2+\lambda \psi(\mathbf{H}),\label{eq10}\\
	&=\mathcal{P}_{\lambda,\psi}\left(\mathbf{V}_{t}\right),\label{eq11}
\end{align}
where $\alpha>0$ is the step size in gradient descent and $\mathcal{P}_{\lambda,\psi}$ is a proximal operator that depends on the regularization function $\psi(\mathbf{H})$. 
Most of the research in mmWave channel estimation literature typically employs the $\ell_1$ norm as a regularization function to promote sparsity in the beamspace channel.
However, using such a hand-crafted regularization function to express the true prior information of the channel can be overly simplistic, which could lead to degraded estimation performance due to the inherent complexity of the channel.
To address this issue, we unfold the PGD algorithm into a deep neural network with a data-driven proximal operator, enabling the network to extract relevant prior information from the channel and perform implicit proximal operations.

The architecture of our proposed TPGD-Net is shown in Fig. 2.
The network comprises $T$ cascaded layers, where each layer corresponds to an iteration of the PGD algorithm.
Each layer consists of a gradient descent module (GDM) and a proximal mapping module (PMM).
The GDM corresponds to the gradient descent step (Eq. (\ref{eq9})) in the PGD algorithm.
The step size $\alpha$ is set as a trainable parameter $\alpha_{t}$ in each layer to enhance  the flexibility of the network and speed up convergence.
The output of the GDM in the $t$-th layer, with input $\mathbf{H}_{t-1}$, is defined as follows:
\begin{equation}
	\mathbf{V}_{t}=\mathbf{H}_{t-1}+\alpha_{t} \mathbf{W}^{{H}}\left(\mathbf{Y}-\mathbf{W H}_{t-1}\right).
\end{equation}
\begin{figure*}[htbp]
	{\includegraphics[scale=0.80]{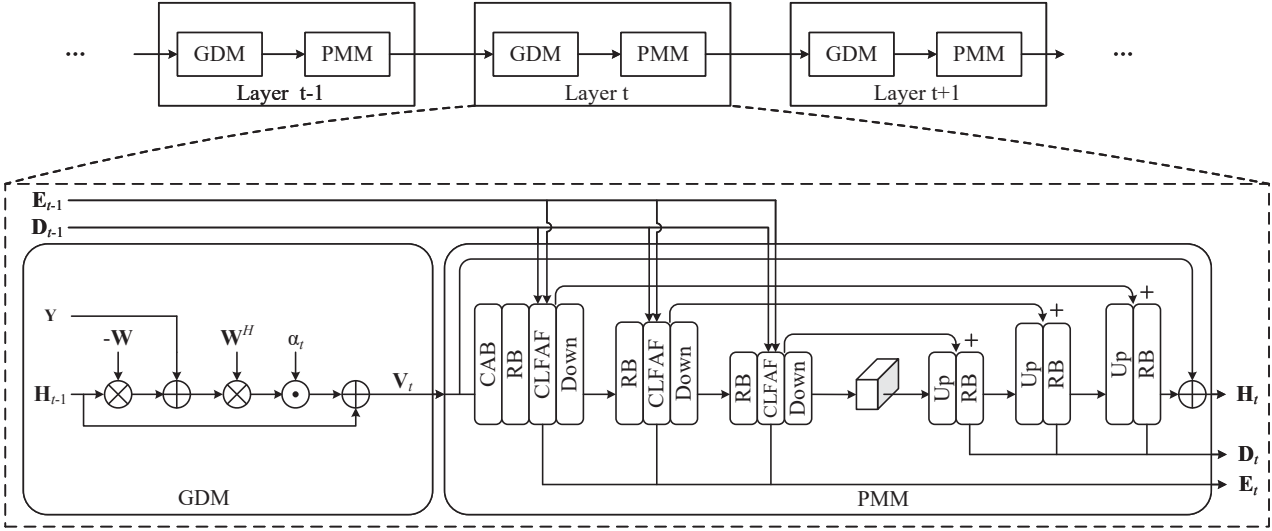}}
	\caption{Architecture of the proposed Trainable Proximal Gradient Descent Network.}
\end{figure*}
The PMM corresponds to the proximal mapping step (Eq. (\ref{eq11})) in the PGD algorithm, which aims to estimate the channel $\mathbf{H}_{t}$ from the input $\mathbf{V}_{t}$.
As demonstrated in Eq. (\ref{eq10}) and Eq. (\ref{eq11}), the regularization function incorporates prior information to obtain the estimated channel $\mathbf{H}_{t}$ and appears only in proximal operator.
Hence, instead of learning a regularization function and explicitly computing its proximal operator, we propose to replace the proximal operator with a neural network.

The proposed neural network, shown in Fig 2, is a U-net with encoder-decoder blocks and skip connections. It employs a channel attention block (CAB) \cite{zamir2021multi} to extract essential features of the beamspace channel and an encoder-decoder block for downsampling and upsampling to obtain multi-scale feature maps. The residual block (RB) \cite{zamir2021multi} is used to extract features at each scale. Max-pooling and bilinear interpolation are used for downsampling and upsampling, respectively. 
Therefore, the proposed neural network bypasses explicit handcrafted regularization function, explores the prior information inherent in mmWave channels, and performs proximal operation implicitly.

Since the PGD algorithm produces estimated channels at each iteration, the TPGD-Net is designed to take these channels as inputs and output estimated channels at each layer.
However, the limited transmission of feature information between layers in TPGD-Net results in reduced channel estimation performance.
To address this limitation, and taking into account the characteristics mmWave channel, which is characterized by sparsity and concentration, we propose a cross-layer feature attention fusion module (CLFAF) based on spatial attention mechanisms\cite{chen2020dasnet}

In particular, the features extracted from the $t$-th layer at the $j$-th scale in the encoder and decoder are denoted as $\mathbf{E}_t^j$ and $\mathbf{D}_t^j$, respectively.
The CLFAF module first concatenates the feature maps $\mathbf{E}_{t}^j$, $\mathbf{D}_{t}^j$, and $\mathbf{E}_{t+1}^j$ along the channel dimension to obtain a feature map of size $3C \times H \times W$, where $C$, $H$ and $W$ are the number of channels, height and width of the feature map respectively.
This concatenated feature map is then fed into a $1 \times 1$ convolution layer to obtain the intermediate feature map $\mathbf{I}_{t+1}^{j}$ of size $C \times H \times W$.
The intermediate feature map $\mathbf{I}_{t+1}^{j}$ is further processed by three convolution layers to obtain feature maps $\mathbf{B}_{t+1}^{j}$, $\mathbf{C}_{t+1}^{j}$, and $\mathbf{D}_{t+1}^{j}$ of size $C \times H \times W$, which are then reshaped to matrices of size $C \times HW$.
The transpose of the matrix $\mathbf{B}_{t+1}^{j}$ is multiplied with the matrix $\mathbf{C}_{t+1}^{j}$, and the resulting matrix is passed through a softmax function to obtain the spatial attention map $\mathbf{S}_{t+1}^{j}$.
Additionally, the transpose of the matrix $\mathbf{S}_{t+1}^{j}$ is multiplied with the matrix $\mathbf{D}_{t+1}^{j}$, and the resulting matrix is then scaled by a learnable parameter $\beta_{t+1}^{j}$ and reshaped to size $C \times H\times W$.
This reshaped matrix is element-wise summed with the feature map $\mathbf{E}_{t+1}^{j}$ to obtain the final feature fusion output $\hat{\mathbf{E}}_{t+1}^{j}$:
\begin{equation}
\hat{\mathbf{E}}_{t+1}^{j}=\beta_{t+1}^{j} (\text{reshape}(({\mathbf{S}_{t+1}^{j}})^T \mathbf{D}_{t+1}^{j})) + \mathbf{E}_{t+1}^{j}.
\end{equation}
The resulting feature $\hat{\mathbf{E}}_{t+1}^{j}$ is a selective fusion of features from the current layer and previous layer, achieved through a spatial attention mechanism that takes into account the spatial relationships among features.
Hence, the CLFAF module improves the transmission of feature information across layers, leading to enhanced channel estimation performance. 
Finally, the output of proximal mapping module can be expressed as:
\begin{equation}
	\mathbf{H}_{t}= \mathcal{P}_{\boldsymbol{\theta}_{t}}\left(\mathbf{V}_{t},\left[\mathbf{E}_{t-1}^{j},\mathbf{D}_{t-1}^{j}\right]_{j=1}^{J}\right),
\end{equation}
where $\boldsymbol{\theta}_{t}$ denotes the parameters of PMM in the $t$-th layer. 
The TPGD-Net can be trained by optimizing its network parameters $\Omega=\left\{\alpha_{t},{\boldsymbol{\theta}_{t}}\right\}_{t=1}^T$ using the stochastic gradient descent algorithm to minimize the loss function:
\begin{equation}
	\mathcal{L}(\Omega)=\sum_{t=1}^T\left\|\mathbf{H}-\mathbf{H}_{t}\right\|_2^2.
\end{equation}

\section{Experiments}
In this section, we evaluate the channel estimation performance of the proposed TPGD-Net through computational simulations.
We consider a scenario where a BS equipped with a lens antenna array of $N = 128$ elements and using $N_{RF}=64$ RF chains serves $K = 8$ single-antenna users using a pilot sequence of length $M = 8$.
The beam selection network uses low-cost one-bit phase shifters, and the elements of ${\mathbf{W}}$ are randomly generated from the set ${\pm {1}/{\sqrt{M N_{R F}}}}$ with equal probability.
The SV channel model in Eq. (\ref{eq1}) and the DeepMIMO dataset are used to generate spatial channels.
For the SV channel model, channels for each user are generated with the following parameters: 1) $L_k=3$ path components; 2) $\beta_{k}^{l} \sim \mathcal{C N}(0,1)$.
For the DeepMIMO dataset, channels for each user are generated using ray-tracing under the indoor scenario 'L2' with the following parameters:  1) number of paths are 5; 2)system bandwidth is 0.02GHz; 3) operating frequency is 28 GHz.
The generated channel data are used to create 5 different signal-to-noise ratios (SNR) ranging from 0 dB to 20 dB, resulting in $5\times 10^7$ data pairs ($\mathbf{Y}$, $\mathbf{H}$) per SNR.
These data pairs are then divided into training, validation, and testing datasets, with $3.5\times 10^7$, $0.5\times 10^7$, and $1.0\times 10^7$ data pairs ($\mathbf{Y}$, $\mathbf{H}$) per SNR, respectively. 

The proposed TPGD-Net utilizes a weight sharing approach to reduce the number of network parameters, where all layers share the same network parameters, except for the first and last layer. 
The number of network layers $T$ is set to $5$, with three scales in both the encoder and decoder in the PMM module. 
The batch size is set to $16$, and the Adam optimizer with a learning rate of $1.0 \times 10^{-4}$ is used for training.
The performance of the trained networks is evaluated on the testing datasets in terms of Normalized Mean Squared Error (NMSE):
\begin{equation}
	\mathrm{NMSE}=\mathbb{E}\left\{\left\|{\mathbf{H}}_{T}-\mathbf{H}\right\|_2^2 /\|\mathbf{H}\|_2^2\right\}.
\end{equation}

The effectiveness of the proposed TPGD-Net is further evaluated by comparing its performance with three state-of-the-art compressed sensing-based channel estimation methods: the AMP algorithm \cite{donoho2009message}, the GM-LAMP network \cite{wei2020deep}, and the the iterative shrinkage-thresholding algorithm (ISTA) \cite{beck2009fast}.
It should be noted that the ISTA \cite{beck2009fast} is a special case of the PGD algorithm that employs the $\ell_1$ norm as the regularization function in Eq. (\ref{eq8}).
Additionally, we also evaluate the performance of the deep generalized unfolding network (DGU-Net) \cite{mou2022deep}, which was originally proposed for image restoration, but we adapt it for channel estimation.
Fig. 3 presents a comparison of the performance of the above-mentioned methods in terms of NMSE versus SNR under the DeepMIMO chhannel dataset.
The GM-LAMP, DGU-Net, and TPGD-Net outperform the AMP and ISTA algorithms due to their ability to learn from data.
Additionally, the DGU-Net and TPGD-Net outperform the GM-LAMP, with DGU-Net achieving a 1.6 dB NMSE gain and TPGD-Net achieving a 2.3 dB NMSE gain at SNR = 10 dB over GM-LAMP. This is due to the ability of the proximal mapping module in both networks to capture more informative features from mmWave channel data and learn a tailored proximal operation.
Moreover, the TPGD-Net outperforms the DGU-Net by approximately 0.8 dB in NMSE at SNR = 10 dB. This improvement is attributed to the CLFAF module in TPGD-Net, which effectively fuses features across layers to provide better features for channel estimation.
\begin{figure}[!t]
\centering
\includegraphics[scale=0.85]{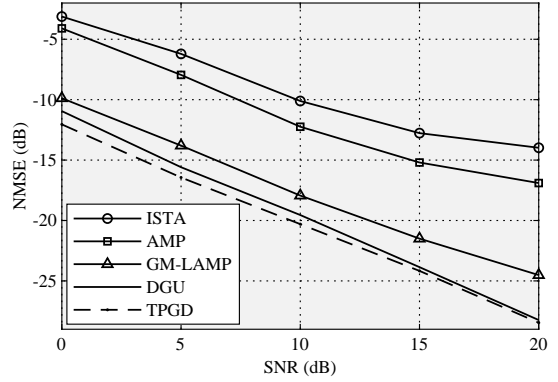}
\caption{NMSE comparison versus SNR under theDeepMIMO channel dataset.}
\label{fig_1}
\end{figure}

The above findings are also supported by the results presented in Fig. 4, where the estimation performance is compared under the SV chhannel model.
To evaluate the robustness of TPGD-Net to SNR, we train TPGD-Net on a training set with SNRs ranging from 2 dB to 22 dB and evaluate it on a test set with SNRs ranging from 0 dB to 20 dB, denoted as TPGD-R-Net.  
The performance of TPGD-R-Net is comparable to that of TPGD-Net, demonstrating the robustness of the proposed TPGD-Net to shifts in noise levels. 
Additionally, we conducte an ablation study by removing the CLFAF module from TPGD-Net, resulting in TPGD-A-Net.
The absence of the CLFAF module in TPGD-Net results in performance degradation.
Specifically, TPGD-A-Net shows a degradation of 2.3 dB compared to TPGD-Net at SNR = 5dB, highlighting the importance of the CLFAF module in improving network performance.

\begin{figure}[!t]
	\centering
	\includegraphics[scale=0.85]{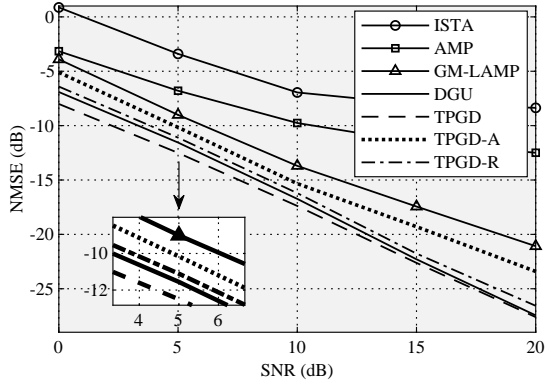}
	\caption{NMSE comparison versus SNR under the SV channel model.}
	\label{fig_2}
\end{figure}


Fig. 5 shows the NMSE performance of TPGD-Net with varying numbers of RF chains, namely $N_{RF}=24, 32$, and $64$, compared with GM-LAMP with $N_{RF}=64$.
The simulation results show that the channel estimation performance of TPGD-Net improves as the number of RF chains at the BS increases. 
In addition, TPGD-Net with $N_{RF}=32$ outperforms GM-LAMP with $N_{RF}=64$, achieving an NMSE gain of approximately 0.93 dB at SNR=5 dB.
Hence, we can conclude that TPGD-Net reduces the number of required RF chains by half compared to GM-LAMP while providing comparable or better performance.

\begin{figure}[!t]
	\centering
	\includegraphics[scale=0.85]{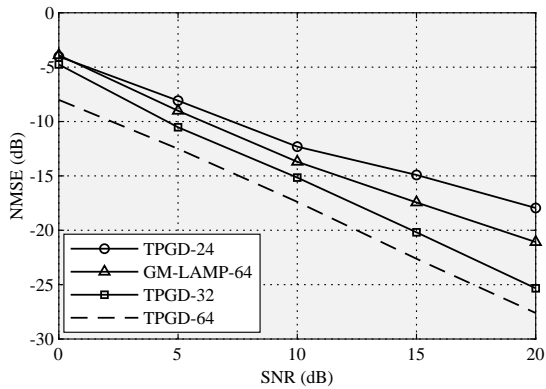}
	\caption{NMSE comparison versus SNR under SV channel model with different number of RF chains.}
	\label{fig_3}
\end{figure}
\section{Conclusion} 
In this letter, we proposed TPGD-Net, a deep learning-based approach for millimetre-Wave channel estimation in massive MIMO communication systems.
The proposed TPGD-Net employs a deep unfolding strategy to transform the proximal gradient descent algorithm into a layer-wise network, with a learnable gradient descent module and a learnable proximal mapping module in each layer.
In addition, we introduced a cross-layer feature attention fusion module to enhance the transfer of feature information across layers.
Simulation results on the Saleh-Valenzuela channel model and the DeepMIMO dataset demonstrate that TPGD-Net outperforms state-of-the-art channel estimators in terms of accuracy while significantly reducing the number of required RF chains.

\bibliographystyle{IEEEtran}
\bibliography{references}{}


 


\vspace{11pt}


\vfill

\end{document}